\documentclass[aps,
        pre,
        reprint,
        superscriptaddress,
        nolongbibliography,
        nofootinbib,
        floatfix,
        notitlepage
        ]{revtex4-2}
\usepackage{amsmath,amssymb,amsfonts}

\makeatletter
\newcommand{\dotr}[1]{%
  \mathpalette\@dotr{#1}%
}
\newcommand*{\@dotr}[2]{%
  \sbox0{$\m@th#1#2$}%
  \usebox{0}%
  \raisebox{\dimexpr\ht0-\height}{$\m@th#1\@smallbullet#1\bullet$}%
  \kern\scriptspace
}
\newcommand*{\@smallbullet}[2]{%
  \scalebox{.5}{$\m@th#1#2$}%
}
\makeatother

\usepackage{natbib}
\usepackage[latin1]{inputenc}

\usepackage[dvipsnames]{xcolor}

\usepackage{hyperref}
\hypersetup{colorlinks=true,citecolor=Red,urlcolor=Blue}

\usepackage{mathrsfs}
\usepackage{bbm}
\usepackage{slashed}
\DeclareSymbolFont{rsfs}{U}{rsfs}{m}{n}
\DeclareSymbolFontAlphabet{\mathrsfs}{rsfs}
\usepackage[mathscr]{eucal}
\usepackage[normalem]{ulem}
\usepackage{verbatim}

\usepackage{graphicx}

\newcommand{\be}{\begin{equation}}
\newcommand{\ee}{\end{equation}}
\newcommand{\bi}{\begin{itemize}}
\newcommand{\ei}{\end{itemize}}
\newcommand{\bea}{\begin{eqnarray}}
\newcommand{\eea}{\end{eqnarray}}
\newcommand{\ud}{\mathrm{d}}		

\newcommand{\LCm}{{\scriptscriptstyle -}}
\newcommand{\LCp}{{\scriptscriptstyle +}}
\newcommand{\LCpm}{{\scriptscriptstyle \pm}}

\newcommand{\LCperp}{{\scriptscriptstyle \perp}}

\definecolor{FG}{RGB}{20,140,20}

\usepackage{microtype}
\usepackage{siunitx}
\DeclareMathOperator{\sech}{sech}
\DeclareMathOperator{\erf}{erf}

\usepackage[caption=false,labelformat=simple,listofformat=subsimple]{subfig}

\begin{document}

\title{Exact solutions in radiation reaction and the radiation-free direction}
\author{Robin Ekman}
\email{robin.ekman@plymouth.ac.uk}
\author{Tom Heinzl}
\email{theinzl@plymouth.ac.uk}
\author{Anton Ilderton}
\email{anton.ilderton@plymouth.ac.uk}
\affiliation{Centre for Mathematical Sciences, University of Plymouth, Plymouth, PL4 8AA, UK}
\begin{abstract}
    We present new exact solutions of the Landau-Lifshitz and higher-order Landau-Lifshitz equations describing particle motion, with radiation reaction, in intense electromagnetic fields.
    Through these solutions and others we compare the phenomenological predictions of different equations in the context of the conjectured `radiation-free direction' (RFD).
    We confirm analytically in several cases that particle orbits predicted by the Landau-Lifshitz equation indeed approach the RFD at extreme intensities, and give time-resolved signals of this behaviour in radiation spectra.
\end{abstract}
\maketitle

\section{Introduction}
Despite having been studied for more than a century~\cite{abraham1905,lorentz1909,dirac1938classical}, radiation reaction (RR) continues to attract theoretical~\cite{DiPiazza:2009zz,bulanov2011lorentz,Iso:2012gn,Zhang:2013ria,Vranic:2014cba,Dinu:2015aci}, computational~\cite{Gonoskov:2014mda,li2020accurately}, and experimental~\cite{Cole:2017zca,Poder:2018ifi,baird2019realising,Nielsen:2020ufz} interest.
In large part, this attention is driven by intense laser systems~\cite{ELI_SDRb,danson2019petawatt,Abramowicz:2021zja,Meuren:2020nbw} now granting access to regimes where quantum and classical RR forces can dominate the Lorentz force.
For recent reviews see Refs.~\cite{Burton:2014wsa,Blackburn:2019rfv}.

It is well-known that including RR effects allows for new phenomena, such as anomalous particle trapping~\cite{Gonoskov:2013aoa}, chaotic motion~\cite{bulanov2017charged}, symmetry breaking~\cite{PhysRevE.84.046409,Harvey:2011dp}, and significantly enhanced generation of certain plasma wave modes~\cite{Gelfer:2018rfy}.
In such scenarios it is often expected that classical RR effects receive significant quantum corrections; the simpler setting of classical physics can, nevertheless, still provide important insight~\cite{Heinzl:2021mji}, and classical effects can persist in quantum theory~\cite{Gonoskov:2013aoa}.
From a theoretical point of view, classical radiation reaction also remains an interesting test-bed for the emergence of non-perturbative physics~\cite{dirac1938classical,RevModPhys.33.37,Zhang:2013ria,Heinzl:2021mji}.

Eliminating the radiation fields, created by the charges, from the classical equations of motion, one arrives at the Lorentz-Abraham-Dirac (LAD) equation~\cite{abraham1905,lorentz1909,dirac1938classical} which contains the third time derivative of position.
This implies unwanted effects such as runaway solutions and pre-acceleration.
The Landau-Lifshitz~\cite{LandauLifshitzII} (LL) equation, obtained from LAD through `reduction of order' (see below) is however free of these difficulties and is typically an excellent approximation below scales where quantum effects appear~\cite{bulanov2011lorentz}.

Given the subtlety of unphysical non-perturbative effects, and the number of complex phenomena attributable to RR, exact solutions can help in making precise statements.
In this paper we present two new exact solutions for field configurations that depend only on a single lightlike direction, with polarisation either longitudinal or transverse to that direction.
Specifically, for the longitudinal case we give the exact solution of the LL equation; and
for transverse polarisation, i.e.~plane waves, we solve the \emph{second-order} Landau-Lifshitz equation obtained by iteration of reduction of order in closed form, in a physically motivated setup.

As an application of these exact solutions (and others to be discussed), we provide analytic evidence confirming the tendency of radiation reaction in strong fields to align particle motion with the `radiation free direction' (RFD) in which they locally experience zero acceleration transverse to their direction of motion, thus minimising radiation losses.
The RFD hypothesis is supported by several analytical results, special cases and numerical simulations~\cite{bulanov2011lorentz,Kazinski:2013vga,Gonoskov2018}.

This paper is organised as follows.
We first give notation and conventions, and write down the Lorentz force, LAD and LL equations for reference.
In Sect.~\ref{sec:longitudinal} we solve the LL equation in longitudinally polarised electric fields of arbitrary strength and form, and demonstrate analytically that orbits transition to the RFD.
We also show that RFD dynamics distinguishes between other proposed classical equations of motion.
In Sect.~\ref{sec:plane-wave} we investigate RFD dynamics in the case of plane wave, i.e~transversely polarised backgrounds, for which the solution of the LL equation is already known, and look for signals of (the transition to) RFD dynamics in emission spectra.
In Sect.~\ref{sec:2LL} we solve the `second-order' LL-like equation, obtained from iteration of reduction of order, in plane waves and compare its predictions with those of the standard LL equation.
We conclude in Sect.~\ref{sec:conclusion}.

\subsection{Notation and conventions}
\label{sec:notation}

We use Heaviside-Lorentz units with $c = \hbar = 1$ and employ lightfront coordinates $x^\LCpm := x^0 \pm x^3$ and $x^\perp := (x^1, x^2)$.
Lightfront momentum components $p^\LCpm$ and $p^\LCperp$ are defined analogously.
It is convenient to introduce a lightlike vector $n^\mu$ such that $x^\LCp = n \cdot x$.

In a background field $F_{\mu\nu}$ the LAD equation of particle motion is obtained from the coupled system of the Lorentz force law and Maxwell's equations by integrating out the dynamical electromagnetic fields.
Writing $f \equiv eF/m$, the LAD equation is
\begin{equation}
    \label{DEF:LAD}
    \ddot{x}^\mu = f^{\mu\nu}{\dot x}_\nu + \tau_0\mathcal{P}^{\mu\nu} \dddot{x}_\nu
    \;,
\end{equation}
where a dot is a proper-time derivative, $\tau_0 := e^2/6 \pi m$ is the characteristic timescale for RR, and $\mathcal{P}^{\mu\nu}$ projects orthogonally to ${\dot{x}}^\mu$.
The standard Lorentz force equation for motion in the background only, i.e.~neglecting radiation and RR, is recovered by setting $\tau_0 = 0$.

The LL equation is obtained by substituting~\eqref{DEF:LAD} back into itself to eliminate the third derivative in favour of new, explicitly $f$-dependent terms, and then neglecting terms of order $\tau_0^2$ and larger.
The LL equation thus found is
\begin{equation}
    \ddot{x}^\mu
    =
    f^{\mu\nu} \dot{x}_\nu + \tau_0
    \dot{f}^{\mu\nu} \dot{x}_\nu
    + \tau_0 \mathcal{P}^{\mu\nu} f_{\nu \rho} f^{\rho \sigma} \dot{x}_\sigma
    \;.
    \label{eq:LL}
\end{equation}
This process is called `reduction of order', since it replaces a third-order ODE, the LAD equation with a second-order ODE.
The process can of course be extended to higher orders in $\tau_0$; we return to this in Sect.~\ref{sec:2LL}.

\section{Longitudinal polarisation}
\label{sec:longitudinal}
We consider first electric fields depending only on $x^\LCp$, which represent electromagnetic pulses propagating in the negative $z$ direction.
We take the fields to be `longitudinally' polarised in the $z$-direction~\cite{Tomaras:2000ag,Woodard:2001hi}.
This is not a solution of the source-free Maxwell equations, but waves of this type can be realised in a plasma or using binary optics for light~\cite{Wang:2008}.
The non-zero components of the field tensor are
\begin{equation}
    \label{eq:LONGFIELD}
    F_{\LCp \LCm} = -F_{\LCm \LCp} = \partial_\LCp A_\LCm(x^\LCp) \;,
\end{equation}
equivalent to having a purely electric field  $\mathbf{E} = (F_{\LCm \LCp}/2)\, \hat{\mathbf{z}}$.
This class includes of course constant electric fields, on which we comment below.
Recall that any massive particle orbit can be parameterised by lightfront time~\cite{Dirac:1949cp}.
Using this, the solution of the Lorentz force equation in our background is easily found; writing $a(x^\LCp) = eA^\LCp$, the first integrals, i.e.~the particle momenta $\pi^\mu= m \dot{x}^\mu$, are
\begin{align}
    \text{(Lorentz)} \quad \pi^\LCp(x^\LCp)       & = p^\LCp - a(x^\LCp) \;,
    \label{eq:Lorentz-solution-p}
    \\
    \label{eq:Lorentz-solution-perp}
    \text{(Lorentz)} \quad    \pi^\LCperp(x^\LCp) & = p^\LCperp \;,
    \,
\end{align}
with initial conditions $\pi^\mu = p^\mu$ at some initial time $x^\LCp = x_0^\LCp$ before the field turns on.
$\pi^\LCm$ follows from the mass-shell condition, $\pi^\LCp \pi^\LCm - \pi^\perp \pi^\perp = m^2$, and the momenta determine the orbit, $x^\mu(\tau)$, via quadratures.

Observe that we have uniform motion perpendicular to the pulse direction in~\eqref{eq:Lorentz-solution-perp}, for any $p^\LCperp$, as the `transverse' directions $x^\LCperp$ decouple.
According to~\cite{Gonoskov2018}, though, when RR is included the particle should move toward the RFD which, for the field~\eqref{eq:LONGFIELD}, is parallel or anti-parallel to the electric field polarisation according to the sign of the charge, in other words the negative or positive $z$ direction.
To see the impact of RR effects, we turn to the LL equation.

We consider first the transverse components of~\eqref{eq:LL}:
\begin{equation}
    \begin{split}
        \label{eq:LL-perp}
        \ddot{x}^\LCperp 
    &= - \tau_0 \dot{x}^\LCp \dot{x}^\LCm (2 f_{\LCp \LCm} )^2
    \dot{x}^\LCperp \\
    &= - \tau_0 (1+ \dot{x}^\LCperp \dot{x}^\LCperp)(2 f_{\LCp \LCm} )^2
    \dot{x}^\LCperp \;,
    \end{split}
\end{equation}
using the mass-shell condition in the second line.
The coefficient of $\dot{x}^\LCperp$ in (\ref{eq:LL-perp}) is strictly negative; dotting with $\dot{x}^\LCperp$ this implies  that the time derivative of $|\dot{x}^\LCperp|$ is negative, and hence the transverse velocity is driven monotonically to zero by RR effects.
As a result, the motion becomes confined to the $tz$ plane, which confirms that the particle momentum indeed becomes aligned with the RFD.
A complementary argument is to note that the transverse equation of motion~\eqref{eq:LL-perp} always has the trivial solution $\dot{x}^\perp = 0$; a stability analysis then shows that this solution represents an `attractor', meaning any deviation from $\dot{x}^\perp = 0$ will be killed by virtue of~\eqref{eq:LL-perp}.

Notably, we find that the system on the attractive sub-manifold, $\dot{x}^\perp = 0$, is integrable;
$\dot{x}^\LCm$ becomes trivially determined from the mass-shell condition, $\dot{x}^\LCm = 1 / \dot{x}^\LCp$, and~\eqref{eq:LL} reduces to a single equation for ${\dot x}^\LCp$,
    \begin{equation}
        \label{eq:xddot}
        m\ddot{x}^\LCp(x^\LCp)
        = -\dot{x}^\LCp \partial_\LCp a - \tau_0  \partial^2_\LCp a \;,
    \end{equation}
which can be solved exactly.
Changing independent variable from proper time to lightfront time,~\eqref{eq:xddot} becomes
\begin{equation}
    \partial_\LCp \pi^\LCp = - \partial_\LCp a(x^\LCp) - \frac{\tau_0}{m}  \pi^\LCp \partial^2_\LCp a(x^\LCp) \;,
\end{equation}
which is readily solved to yield, for $\pi^\LCp(-\infty) = p^\LCp$,
\begin{align}
    \pi^\LCp(x^\LCp)    & = e^{- \frac{\tau_0 a'(x^\LCp)}{m}} \Big(
        p^\LCp - \int\limits_{-\infty}^{x^\LCp} e^{ \frac{\tau_0}{m} a'(y)} a'(y) \, \ud y
    \Big)
    \label{eq:LL-solution-p}
    \;.
    \end{align}
As a check, we note that for a constant electric field, $a''=0$, and $p^\LCperp = 0$, the LL solution~\eqref{eq:LL-solution-p} and the Lorentz solution~\eqref{eq:Lorentz-solution-p} agree, and also solve the LAD equation.
(In this case (\ref{eq:LL-perp}) also decouples from the equation for $\pi^\LCp$, and becomes integrable.)
This is consistent with the literature result that there is `no radiation reaction for hyperbolic motion'~\cite{fulton1960classical,Yaremko:2014qoa,Seipt:2019dnn}; there is though radiation~\cite[p.~399]{fulton1960classical,schwinger1998classical}.

With~\eqref{eq:LL-solution-p} one can make explicit the stability of the RFD solution.
Linearising the LL equation in $\pi^\LCperp$, the equations for $\pi^\LCpm$ are unchanged, (since $\pi^\LCperp$ enters them \emph{quadratically}), which allows us to solve~\eqref{eq:LL-perp} as
\begin{equation}
    \pi^\LCperp(x^\LCp)
    \simeq \pi^\LCperp(x^\LCp_0) \exp \big[
        -4 m \tau_0 \int_{x_0^\LCp}^{ x^\LCp } f^2_{\LCp \LCm} / \pi^\LCp \, \ud y
    \big]
    \;,
    \label{eq:LL-solution-perp}
\end{equation}
where $\pi^\LCp$ is as in~\eqref{eq:LL-solution-p}; thus deviations of motion from the RFD are exponentially suppressed.

\subsection{Example: Sauter pulse}
To understand the physics of the exact solution~\eqref{eq:LL-solution-p} in non-constant fields we turn to an explicit example with a chosen field configuration.
We consider a Sauter pulse of width $1/\omega$ and peak field strength determined by~$a_0$,
\begin{align}
    a'(x^\LCp) & = \frac{m \omega a_0}{2} \sech^2(\omega x^\LCp)
    \label{eq:sauter-field}
    \\
    a(x^\LCp)  & = \frac{m a_0}{2} \big(1 + \tanh ( \omega x^\LCp ) \big)
    \,
    ,
    \label{eq:sauter-pot}
\end{align}
with positive/negative values of $a_0$ corresponding to an electric \emph{force} on the particle parallel/anti-parallel to the direction of pulse propagation (which, recall, is the negative $z$-direction in our conventions).

For this field the integral in~\eqref{eq:LL-solution-p} can be performed analytically in terms of the error function with the result, for $a_0>0$,
\begin{widetext}
    \begin{align}
        \pi^\LCp(x^\LCp) =
        e^{ - \frac{\omega \tau_0 a_0}{2} \sech^2 (\omega x^\LCp) } \left[
            p^\LCp - m \sqrt{ \frac{\pi a_0 }{2 \omega \tau_0} } e^{ \frac{\omega \tau_0 a_0}{2} } \left(
                \erf \sqrt{\frac{\omega \tau_0 a_0}{2}} \tanh \omega x^\LCp + \erf \sqrt{\frac{\omega \tau_0 a_0}{2}}
            \right)
        \right]
        \, ,
        \label{eq:LL-sauter-sol}
    \end{align}
\end{widetext}
while for $a_0 < 0$, absolute value signs should be inserted under the square roots and $\erf$ should be replaced by $\operatorname{erfi}$.
The corresponding Lorentz force results are given simply by inserting~\eqref{eq:sauter-pot} into~\eqref{eq:Lorentz-solution-p}.

Note that the solutions of both the LL and Lorentz equations are valid only as long as $\pi^\LCp > 0$, as is required for massive particles.
For $\pi^\LCp \to 0$, the particle is accelerated to almost co-propagate with the field, and reaches the speed of light in finite \emph{lightfront} time, after which the particle `leaves the spacetime manifold'~\cite{Tomaras:2000ag,Woodard:2001hi}.
Note that this finite lightfront time corresponds to \emph{infinite} lab-frame, or proper, time.
As such, considering both~\eqref{eq:Lorentz-solution-p} and~\eqref{eq:LL-solution-p}, it is clear that the sign of $a_0$ is a crucial factor in determining properties of the motion.

\begin{figure}[b!!]
    \centering
    \includegraphics[width=\linewidth]{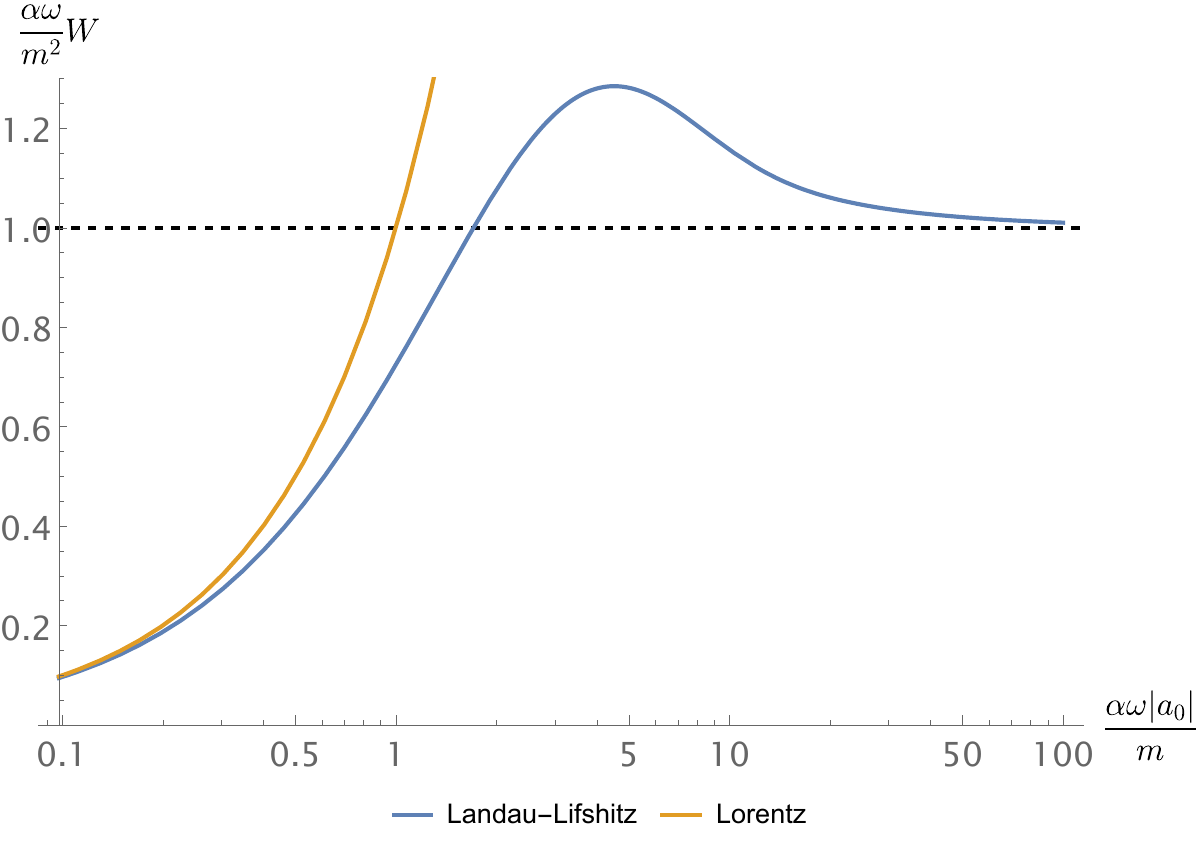}
    \caption{
        Momentum transfer $W$~\eqref{def:W}, in a head-on collision with a Sauter pulse with $a_0<0$.
        Without RR, the momentum transfer is unbounded.
        With RR, $W$ approaches a constant at large $|a_0|$, as RR effects dominate over the electric force.
    \label{fig:acc-pfinal}}
\end{figure}

We consider first the case $a_0<0$, for which the force on the particle is anti-parallel to the propagation direction; for a head-on collision, this means the particle is accelerated in its direction of initial propagation.
We define the momentum transfer $W$ by
\begin{equation}
    \label{def:W}
    W := \pi^\LCp(\infty) - \pi^\LCp(-\infty) = \pi^\LCp(\infty) -p^\LCp
    \;.
\end{equation}
Without RR, $W_{\text{Lorentz}} = m|a_0|$, but according to the LL equation RR reduces the momentum transfer;
one finds from~\eqref{eq:LL-sauter-sol} that, for $\alpha = e^2/(4\pi)$ the fine-structure constant,
\begin{equation}
    W_{\text{LL}}  \simeq
    \frac{m^2}{\alpha \omega}
    \quad
    \text{for}
    \quad
    \frac{\alpha \omega |a_0|}{m} \gg 1 \;.
\end{equation}
The behaviour of $W$ as a function of $a_0$, for both the Lorentz and LL equations of motion, is shown in~Fig.~\ref{fig:acc-pfinal}.

\begin{figure}[t!]
    \centering
    \includegraphics[width=\linewidth]{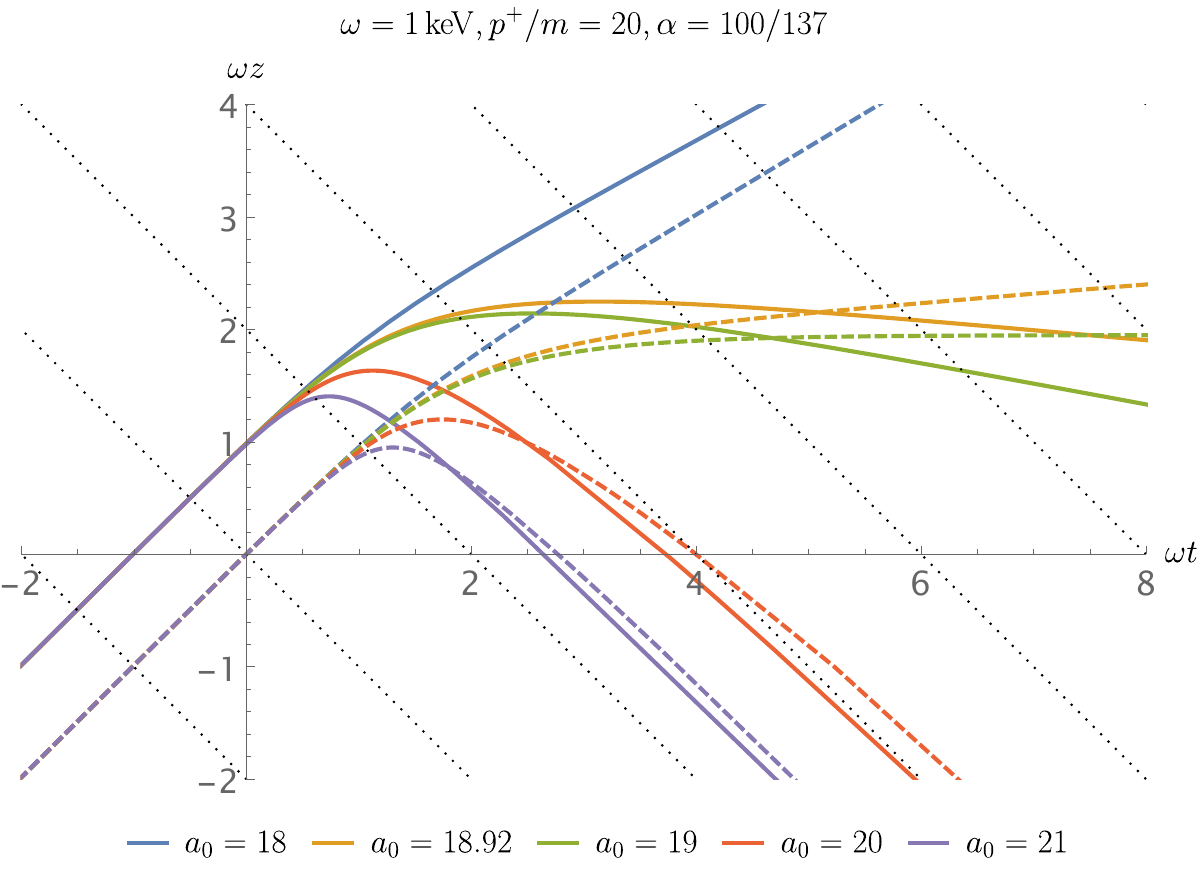}
    \caption{
        Particle orbits in a head-on collsion with a longitudinally polarised Sauter pulse.
        Solid: LL, dashed: Lorentz.
        The LL orbits have been displaced by unity in the $\omega x^\LCm$ direction to distinguish them from the Lorentz orbits.
        Note that an unphysically large $\alpha$ has been used to exaggerate RR.
    }
    \label{fig:ret-orbits}
\end{figure}

We turn to the case $a_0>0$, for which the force on the particle is parallel to the field propagation direction (and therefore opposite the initial propagation direction of the particle).
In this case we find that for sufficiently large $a_0$, the particle can be brought to rest, and for even larger $a_0$ can be caught in the pulse and accelerated almost to the speed of light.
In the Lorentz case, the respective thresholds above which these phenomena occur are easily read off from~\eqref{eq:Lorentz-solution-p} as
\begin{equation}
    a_{0, \text{stop}} = p^\LCp/m - 1
    \quad \text{and} \quad
    a_{0, c} = p^\LCp/m
    \,
    .
\end{equation}
Including RR through the LL equation, the thresholds are the solutions of an intractable transcendental equation, but it is easily found by numerical investigation that both thresholds are lowered by RR.
What this means is that there is a range of $a_0$ such that particles are back-scattered by RR effects, i.e.~their direction of motion is reversed.
Example orbits illustrating this effect are plotted in Fig.~\ref{fig:ret-orbits}.

In the figure,  an unphysically large value of $\alpha$ has been used to exaggerate the effect of RR (along with an $\omega$ much larger than is phenomenologically motivated).
This is because the solution~\eqref{eq:LL-solution-p} describes a particle already moving in the RFD, and because longitudinal forces produce much less radiation than transverse forces, RR is consequently a very small effect.
RR will naturally be more pronounced in the transition to the RFD, i.e.~as $\pi^\LCperp$ is driven to zero and, when close to zero, follows~\eqref{eq:LL-solution-perp}.
Rather than analyse the radiation spectra from this approximate final stage motion, though, we will consider in the next section the case of plane waves, which admit exact solutions to the LL equation for arbitrary initial conditions.
Before doing so we compare the results above with those of other classical equations.

\subsection{Alternative equations}
We ask here what different classical equations predict regarding transition to, and motion in, the RFD.
The question is somewhat academic, as unlike LAD (the classical equation of motion) or LL (an approximation to it) other equations proposed over the years bring in some external assumption.
We will therefore be brief.

Eliezer's equation~\cite{Eliezer} (rederived by Ford-O'Connell in a slightly different context~\cite{FOC}) is
\begin{align}
\ddot{x}^\mu
    & =
    f^{\mu\nu} \dot{x}_\nu + \tau_0 \,  \mathcal{P}^{\mu\nu} \, \frac{\ud}{\ud\tau} (f_{\nu\rho} \dot{x}^\rho) \; .
        \label{eq:FOC}
\end{align}
Linearising in the transverse momentum as above, we find
\be\label{FOCsol}
    {\dot \pi}^\LCperp \simeq \bigg(\frac{\tau_0}{m} \frac{{\dot \pi}^\LCp}{\pi^\LCp} \partial_\LCp a \Bigg)\pi^\LCperp \;, \qquad \pi^\LCp = \pi^\LCp_\text{LL} \;,
\ee
This means that motion \emph{in} the RFD, when it is reached, is exactly as described by LL.
The approach to the RFD is, however, not monotonic: numerical investigation confirms that the coefficient of ${ \pi}^\LCperp$ in~\eqref{FOCsol} is initially negative, but can in principle change sign, implying attraction to or repulsion from the RFD (though this tends to happen only for rather extreme parameters).

Ultimately, it is impossible to turn off quantum effects, and the predictions of some proposed equations do not agree with the classical limit of QED results~\cite{Krivitsky:1991vt,Higuchi:2002qc,Ilderton:2013dba}.
An example is the Mo-Papas equation~\cite{MP},
\begin{align}
    \ddot{x}^\mu
    & = f^{\mu\nu} \dot{x}_\nu + \tau_0 \,
    \mathcal{P}^{\mu\nu} f_{\nu\rho} \ddot{x}^\rho
    \; .
    \label{eq:MP}
\end{align}
Linearising again, one finds in this case that
\begin{equation}
    {\dot{\pi}}^\LCperp \simeq -\frac{\tau_0}{m^2}(\partial_\LCp a)^2 \pi^\LCperp
    \;,
    \quad
    \pi^\LCp = \pi^\LCp_\text{Lorentz}
\end{equation}
The first of these equations tells us that, for small transverse momentum, ${\dot \pi}^\LCperp =0$ is again a stable attractor, and the orbit goes to the RFD.
However, motion in the RFD is that predicted by the Lorentz force equation, without radiation reaction.
Thus the LL, Eliezer and Mo-Papas, equations all predict different behaviour, and it seems that, in principle, RFD dynamics can differentiate between classical equations of motion.

\section{Transverse polarisation}
\label{sec:plane-wave}

We turn now to plane waves, i.e.\ functions of $x^\LCp = n \cdot x$, which are transversely, rather than longitudinally polarised.
Their potential $a_\mu:=eA_\mu$ can always be written $a_\mu = m a_0 \delta_\mu^\LCperp f_\LCperp(\omega x^\LCp)$ in which $\omega$ is some frequency scale, $f_\LCperp$ describes the shape of the field, and the dimensionless invariant $a_0$ characterises the peak field strength.

\begin{figure}[b!]
    \centering
    \includegraphics[width=\linewidth]{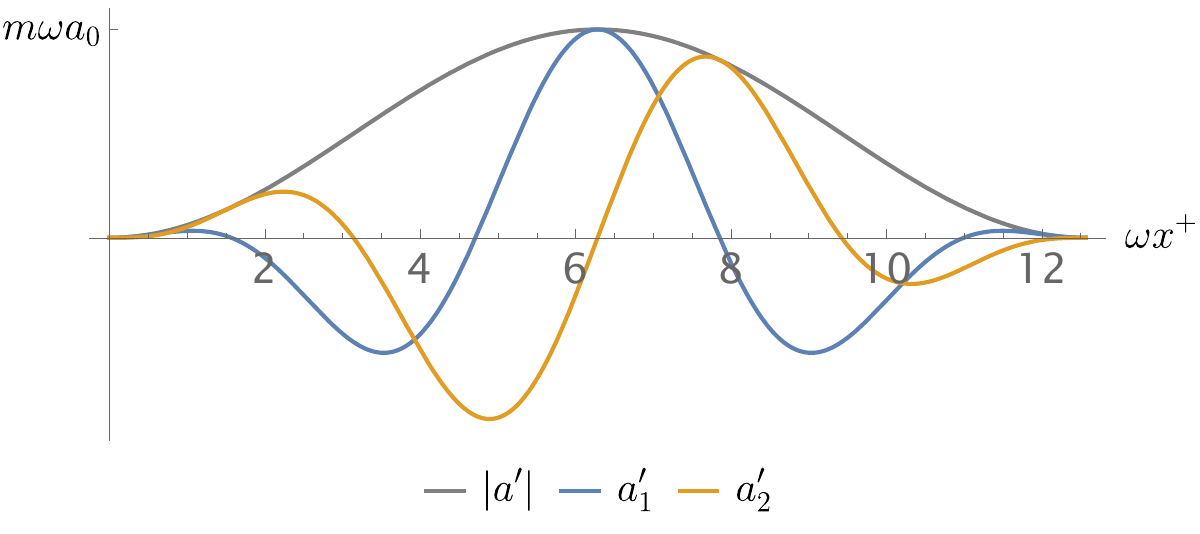}
    \caption{Electric field of a circularly polarised pulse with a $\sin^2$ envelope, cf.~\eqref{eq:sin-sq-pulse}.}
    \label{fig:sin-sq-pulse}
\end{figure}

The solution to the LL equation was first given for monochromatic, linearly polarised plane waves in Ref.~\cite{Heintzmann:1972mn}, and extended to all plane waves in Ref.~\cite{PiazzaExact}.
It is conveniently presented using the following parameterisation of the momentum $\pi_\mu$,
\begin{equation}
    \pi^\mu
    = \frac{1}{h} \bigg( p^\mu - B^\mu + n^\mu \frac{2 p \cdot B - B^2 + m^2 (h^2 - 1)}{2n \cdot p} \bigg)\;,
    \label{eq:plane-wave-sol}
\end{equation}
in which $B_\mu$ is purely transverse.
Writing $\phi\equiv \omega x^\LCp$, $k_\mu \equiv \omega n_\mu$ and a prime for a $\phi$-derivative, the functions $h\equiv h(\phi)$ and $B_\mu\equiv B_\mu(\phi)$ are
\begin{align}
    h(\phi) & = 1 - \tau_0  \frac{k \cdot p}{m^3} \int_{-\infty}^{\phi}\! \ud y \, a'(y) \cdot a'(y) \;,
    \label{eq:h-def}
    \\
    B_\mu(\phi) &=
    \tau_0 \frac{k \cdot p}{m} a'_\mu(\phi)
    + \int_{-\infty}^{\phi} \ud y \, h(y) a_\mu'(y) \;,
    \label{eq:B-def}
\end{align}
with initial conditions $\pi^\mu(-\infty) = p^\mu$.
The parameterisation~\eqref{eq:plane-wave-sol} is chosen because it makes clear that
\emph{all} momentum components are actually proportional to $1/h =\pi^\LCp /p^\LCp$; in this sense~$\pi^\LCp$ governs the dynamics.

It is more convenient to analyse the transition to the RFD at the level of the solution~\eqref{eq:plane-wave-sol}, rather than at the level of the LL equation, as we did for longitudinally polarised fields.
First, the asymptotic scaling of the outgoing momentum components with pulse length $T$ and intensity $a_0$ can be found using~\eqref{eq:h-def}--\eqref{eq:B-def}. We have that $h \sim T a_0^2, B \sim T^2 a_0^3$, which implies
\begin{equation}
    \pi^\LCp(\infty) \sim 1/(T a_0^2)
    \quad
    \pi^\LCperp(\infty) \sim T a_0
    \quad
    \pi^\LCm(\infty) \sim T^3 a_0^4
    \,
    .
\end{equation}
From this we see that the ratio of transverse momentum to $z$-momentum,
\begin{equation}
    \frac{|\pi^\LCperp|}{\pi^z} = \frac{2|\pi^\LCperp|}{\pi^\LCp-\pi^\LCm}
    \;,
    \label{eq:momentum-ratio}
\end{equation}
behaves as $\sim T^{-2} a_0^{-3}$ for $a_0\gg 1$.
Hence, while motion in a plane wave can never be confined exactly to the $tz$ plane due to the explicitly transverse polarisation, motion is \emph{dominantly} in the $z$ direction for $a_0 \gg 1$.
This is indeed the radiation-free direction according to Ref.~\cite{Gonoskov2018}, although the limits of the formulae provided there need to be taken with care as, in a plane wave, both field invariants vanish.

We illustrate the particle motion using
a circularly polarised, few-cycle pulse with a $\sin^2$ envelope, for which the integrals in~\eqref{eq:h-def} and~\eqref{eq:B-def} can be performed analytically.
The pulse is given by
\begin{equation}
    \begin{pmatrix}
        a_1' \\ a_2'
    \end{pmatrix}
    =
    m a_0
    \sin^2 \frac{\phi}{4}
    \begin{pmatrix}
        \cos \phi \\
        \sin \phi
    \end{pmatrix}
    \label{eq:sin-sq-pulse}
\end{equation}
for $0 \le \phi \le 4 \pi$, and vanishing otherwise, see Fig.~\ref{fig:sin-sq-pulse};
the envelope ensures that the electromagnetic fields, and their derivatives, vanish at the edge of the pulse.

\begin{figure}[tb]
    \centering
    \includegraphics[width=\linewidth]{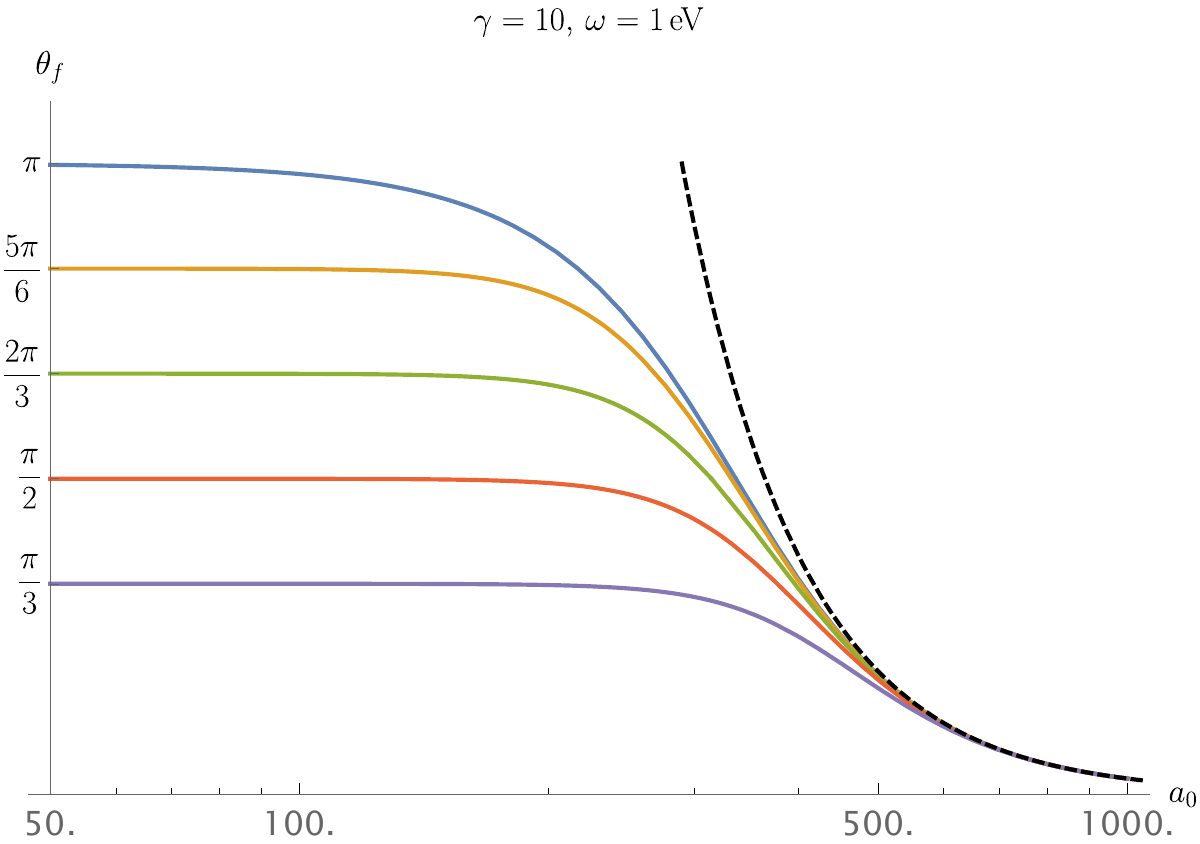}
    \caption{
        Outgoing scattering angle, relative to the laser propagation direction, as a function of $a_0$, at fixed initial particle energy and for several incoming collision angles, indicated at each solid curve, in the pulse~\eqref{eq:sin-sq-pulse}.
        The asymptotic $1/a_0^3$ scaling is also shown (dashed).
        (The scattering angle is essentially independent of $a_0$ below the range displayed.)
        Optical frequency, $\omega = \SI{1}{eV}$, is chosen as being the most phenomenologically relevant; the asymptotic behaviour then sets in for an $a_0$ of several hundred, owing to the smallness of $\omega \tau_0 \simeq 10^{-8}$.
    }
    \label{fig:trans-angle-final}
\end{figure}

Performing the integrals in~\eqref{eq:h-def} and~\eqref{eq:B-def}, we can finesse the approximation~\eqref{eq:momentum-ratio} and calculate the scattering angle~$\theta_f$ of the outgoing electron, taken with respect to the laser propagation direction.
We find
\begin{equation}
    \label{eq:scattering-angle}
    \theta_f \approx \frac{24 m}{7\pi \alpha a_0^3 \omega}
    \qquad
    a_0 \gg 1
    \,
    ,
\end{equation}
which, note, is completely independent of the \emph{incoming} direction of the particle, and is clearly in agreement with~\eqref{eq:momentum-ratio}.
Fig.~\ref{fig:trans-angle-final} shows the full $a_0$-dependence, and the asymptote~\eqref{eq:scattering-angle}, for several incidence angles.

\begin{figure}[bt]
    \centering
    \includegraphics[width=\linewidth]{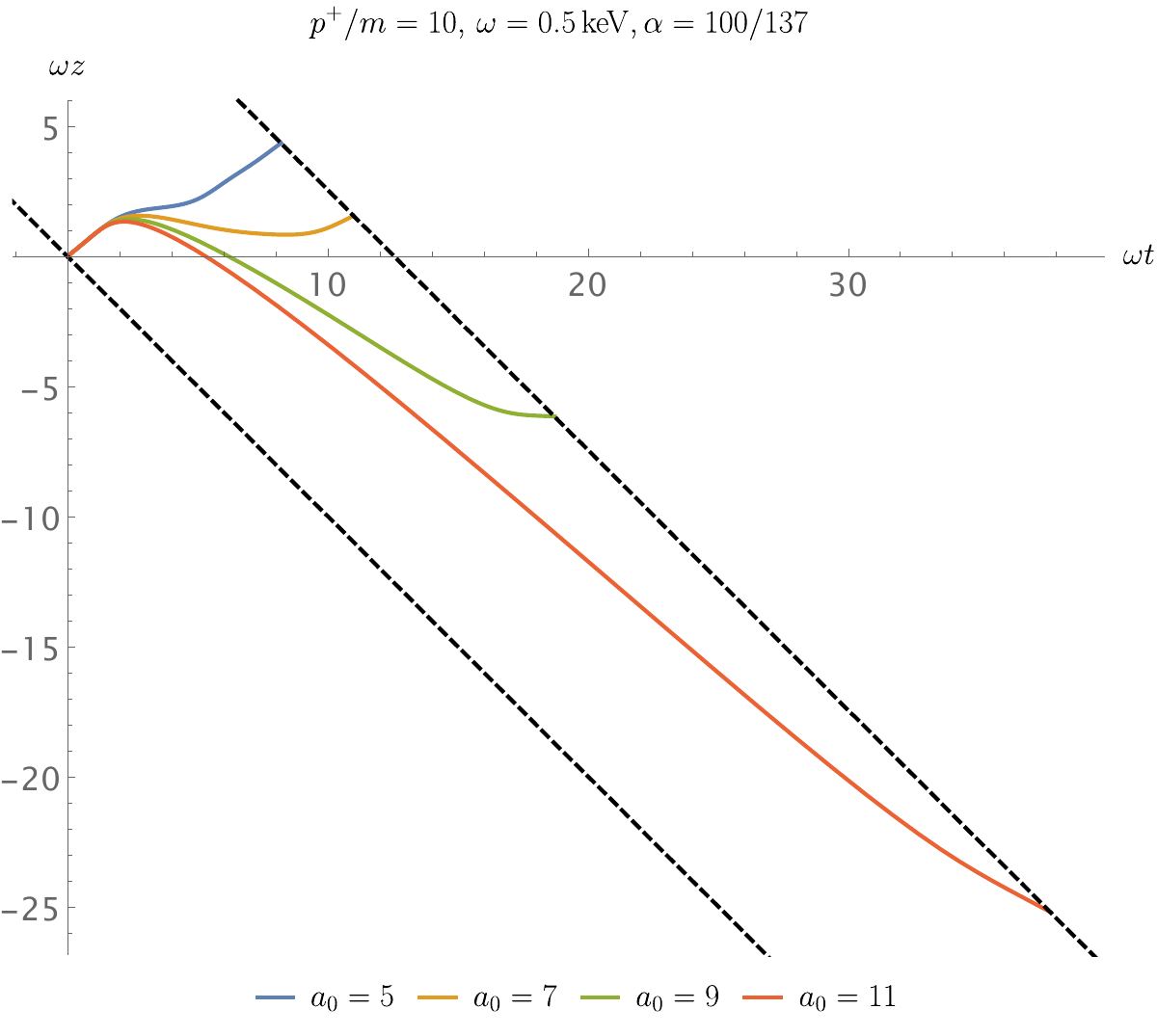}
    \caption{
        Particle orbits for a head-on collision with the pulse~\eqref{eq:sin-sq-pulse}, whose spacetime extent is indicated by the dashed lines.
        As $a_0$ increases, the particle goes from scattering forward, to coming to a stop, to scattering backward.
        (Note: an unphysically large $\alpha$ has been used to exaggerate RR.)
    }
    \label{fig:trans-orbits}
\end{figure}

This confirms that, at high intensity, all particles are driven to propagate dominantly in the RFD (with relatively small transverse momentum).
We illustrate the transition to (near) laser-collinear scattering by plotting particle orbits in the $(t,z)$ plane in Fig.~\ref{fig:trans-orbits}.
Note that the parameters in Fig.~\ref{fig:trans-orbits} have been chosen for visual clarity rather than physicality: for realistic parameters and $a_0$ in the radiation-dominated regime, the particle is carried with the pulse for many cycles's worth of \emph{lab} time.

\subsection{Radiation spectra}

\begin{figure}[bt]
    \centering
    \subfloat[\label{fig:trans-spectrum}]{%
        \includegraphics[width=.98\columnwidth]{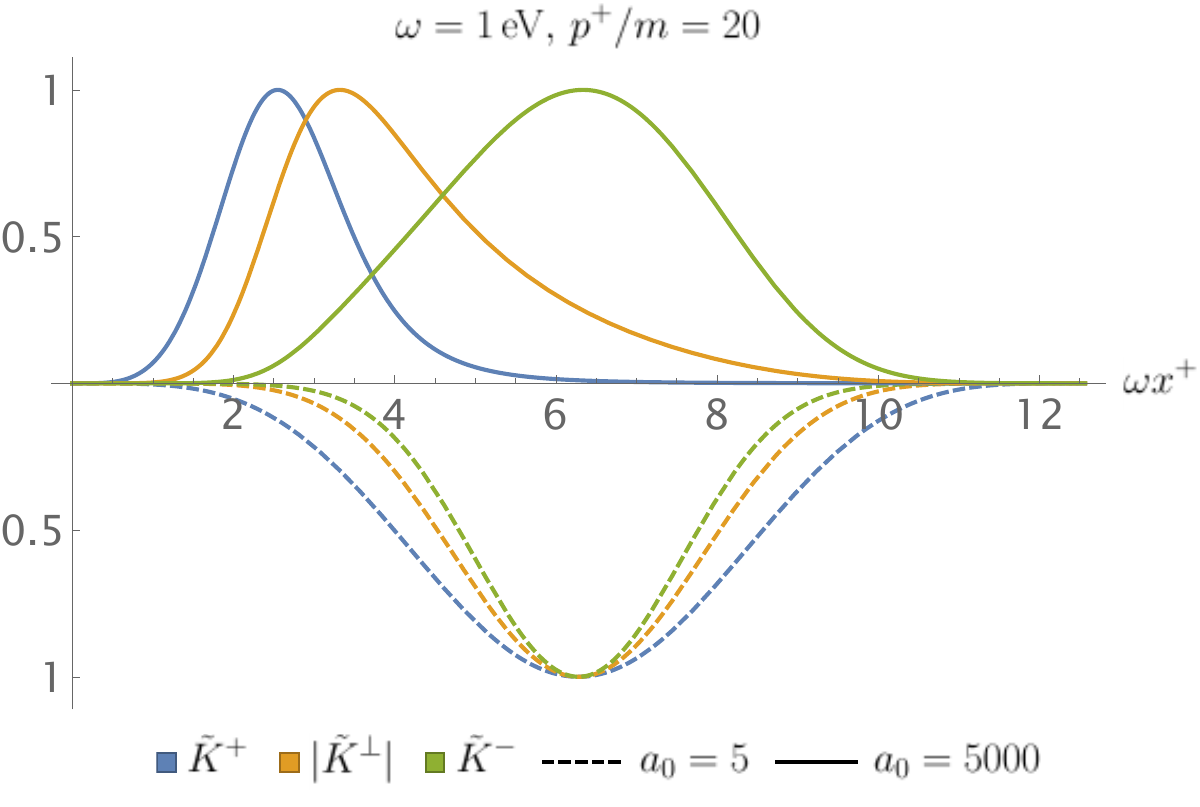}
    }\\
    \subfloat[\label{fig:trans-orbit}]{%
        \includegraphics[width=0.75\columnwidth]{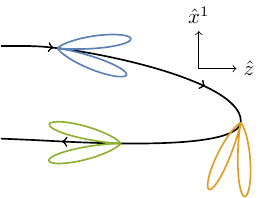}
    }
    \caption{
        Radiation reaction makes the radiation profile in a pulse time-resolved.
        \protect\subref{fig:trans-spectrum}: The radiated 4-momentum in the pulse~\eqref{eq:sin-sq-pulse}, as a function of lightfront time.
        Each component has been normalised to its maximum, $\tilde{K}_\mu = K_\mu/\max |K_\mu|$, and the dashed curves have been reflected in the horizontal axis for clarity.
        \protect\subref{fig:trans-orbit}: Sketch of the particle orbit and the radiation cone rotating along it.
    }
    \label{fig:trans}
\end{figure}

Signatures of RFD dynamics can be found in the emitted radiation spectra.
The radiated photon 4-momentum $K^\mu$ has the standard form
\begin{equation}
        K^\mu = - \int \frac{\ud^2 \ell^\LCperp \ud \ell^\LCp}{(2 \pi)^3 2 \ell^\LCp}\, \ell^\mu |j(\ell)|^2 \;,
        \label{eq:spectrum}
\end{equation}
where $j_\mu$ is the current in Fourier space,
\begin{equation}
    j_\mu(\ell) = - e \int \ud x^\LCp \, e^{i \ell \cdot X(x^\LCp) } \frac{\partial}{\partial x^\LCp} \left( \frac{\pi_\mu}{i \ell \cdot \pi} \right)
    \, ,
\end{equation}
and $X^\mu$ is the particle orbit.
All  but one of the integrals in~\eqref{eq:spectrum} can be performed exactly; the final expression for $K^\mu$ is
\begin{equation}
    K^\mu =
    \frac{2}{3} \frac{e^2}{4\pi} \frac{ k\cdot p}{m^2}
    \int\! \ud \phi \, \frac{m^2 h'(x)^2 - B'(x)^2}{m^2 h(x)^3} \pi^\mu(\phi)
    \;.
\end{equation}
Examining the integrand shows us from where in the pulse radiation is generated.
For low $a_0$ RR is negligible, and we find that emission of all spectral components $K^\mu$ is supported mainly near the peak of the pulse.
A signature of the transition to the RFD is then that emission into different spectral components $K^\mu$ becomes \emph{time-resolved} at high intensity; as shown in  Fig.~\ref{fig:trans-spectrum}, radiation  emitted anti-collinear/transverse/collinear to the laser is predominantly emitted near the beginning/peak/end of the pulse.
This corresponds directly to the change in particle direction associated with transition to the RFD, as sketched in Fig.~\ref{fig:trans-orbit}.

In Fig.~\ref{fig:trans-spectrum} the components are normalised to their maxima to highlight their different temporal supports.
To extract analytic estimates for their relative sizes, when $a_0 \gg 1$, requires a little care.
The components $K^\LCperp$, $K^\LCm$ scale according to naive estimates using~\eqref{eq:plane-wave-sol}, viz.,
\begin{equation}
    K^\LCperp = \mathcal{O}\left( a_0 \right)
    \quad \text{and} \quad
    K^\LCm = \mathcal{O}\left( a_0^4 \right)
    \,
    .
    \label{eq:scaling-others}
\end{equation}
However, the dominant contribution to $K^\LCp$ does not scale as $a_0^{-2}$ as might be expected; the dominant contribution comes from a boundary term hidden in the contribution from ${B'}^2$ and is~\cite{Heinzl:2021mji}
    \begin{equation}
        \frac{K^\LCp}{p^\LCp} \sim -\frac{2}{3} \frac{e^2}{4 \pi} \frac{k \cdot p}{m^4}
        \int\! \ud \phi \, \frac{h^2 {a'}^2}{h^4}
        =
        - \int\! \ud \phi \, \frac{h'}{h^2}
        = 1
        \, .
        \label{eq:scaling-lightfront}
    \end{equation}
This scaling behaviour is a consequence of conservation of total lightfront momentum;
$K^\LCp$ and $\pi^\LCp$ both being non-negative, the radiation field cannot carry off more than the initial $p^\LCp$ of the particle. (The locally constant field approximation to $K^\LCp$ is given in Ref.~\cite{Heinzl:2021mji}, see also Ref.~\cite{DiPiazza:2021nsx}.)

While the total perpendicular momentum is also conserved, $\pi^\LCperp$ is not constrained and can always absorb an arbitrarily large recoil.
Finally, due to the lack of symmetry in $x^\LCp$ the collinear momentum is simply not conserved, putting no restriction on the particle ``pumping'' laser momentum into its own radiation field.

It may seem paradoxical that, comparing~\eqref{eq:scaling-lightfront} and~\eqref{eq:scaling-others}, the dominant component is in the RFD, but we must remember that motion in the RFD minimises radiation \emph{for a given magnitude of the applied force}~\cite{Gonoskov2018}.
In the radiation-dominated regime, though, the particle sheds all of its lightfront momentum early in the rise of the pulse.
As the particle approaches the radiation-free direction the magnitude of the force and, eventually, the Lorentz factor increase, both of which strongly enhance the radiated power.

\section{LL to second order}
\label{sec:2LL}

\newcommand{\LLn}[1]{\ensuremath{\text{LL}_{#1}}}

The LL equation~\eqref{eq:LL} has been obtained by iterating the LAD equation~\eqref{DEF:LAD} to \emph{first} order in $\tau_0$.
To assess the quality of this approximation we consider the size and influence of corrections by performing the iteration to second order in~$\tau_0$.
We will refer to the LAD equation iterated to $n$:th order in $\tau_0$ as LL$_n$ (so that the Lorentz equation of motion is \LLn{0}).
A somewhat lengthy calculation yields the following expression for \LLn{2},
\begin{equation}
    \begin{split}
        \label{2LL-general}
        \ddot{x}^\mu
        = & \big(
            f+\tau_0 \dot f + \tau_0^2 \ddot f -2\tau_0^2 (\dot x f^2 \dot x) f
        \big)^{\mu\nu}{\dot x}_\nu \\
          & + \tau_0\mathcal{P}^{\mu\nu} \big[
              f^2_{\nu\rho} +2 \tau_0 f^3_{\nu\rho} + 2 \tau_0 \dotr{(f^2)}_{\nu\rho}
          \big] {\dot x}^\rho \;.
    \end{split}
\end{equation}

\subsection{Transverse polarisation}
As for \LLn{1}, if one can solve \LLn{2} for $\pi^\LCp$ in a plane wave, as a function of $x^\LCp$, then the transverse momenta $\pi^\LCperp$ are easily calculated, and $\pi^\LCm$ is given by the mass-shell condition.
We therefore focus on $\pi^\LCp$.

The momentum $\pi^\mu$ is again conveniently parameterised as in~\eqref{eq:plane-wave-sol}, so that we again have $\pi^\LCp/p^\LCp = 1/h$.
It is useful to define the two dimensionless RR parameters
\begin{equation}
    \delta = \frac{2}{3}\frac{e^2}{4\pi}\frac{k\cdot p}{m^2} \;,
    \qquad
    \Delta := a_0^2 \delta \;,
\end{equation}
where $\delta$ is essentially an energy parameter, which should strictly be small in the classical regime, whereas $\Delta$ depends on the field strength, and so can be large.
Denoting $\phi$-derivatives with a prime as before, the \LLn{2} equation for $h$ becomes
\begin{equation}
    \label{eq:h-2LL}
    h' = -\frac{\delta}{m^2} a'\cdot a' - \frac{4\delta^2}{m^2 h} a'\cdot a''
    \;,
\end{equation}
which is nonlinear due to the factor of $h^{-1}$ in the final term: this is an Abel equation of the second kind, and thus not analytically solvable in general, while special cases are usually only solvable parametrically~\cite{Polyanin}.

\begin{figure}[t]
    \includegraphics[width=\columnwidth]{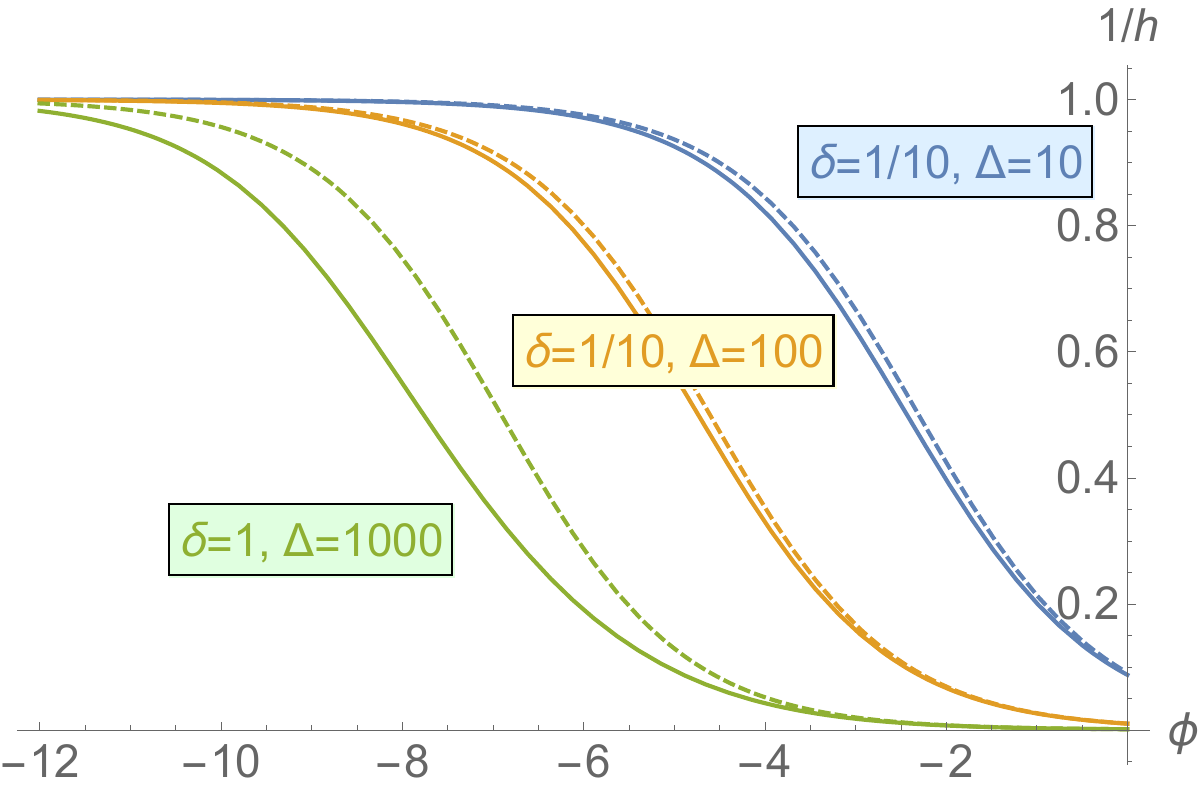}
    \caption{
        \label{FIG:PRODUCTLOG}
        Momentum $\pi^\LCp/p^\LCp \equiv 1/h$ in the rise of the pulse, for various $\delta$ and $\Delta$, in \LLn{2} (solid lines) and \LLn{1} (dashed lines).
        Increasing the field strength, so $\Delta$, leads to significant momentum loss earlier in the pulse.
The rate of this loss is corrected by effects of $\mathcal{O}(\delta)$ in \LLn{2}.}
\end{figure}

However, there is a solvable case which is relevant to the physical situation of interest.
We have seen that at high intensities, a particle can be stopped and turned around soon after it enters the pulse, i.e.~before reaching the peak.
We therefore consider dynamics in the initial \emph{rise} of the pulse, a simple model of which is
\begin{equation}
    \label{rising-def}
    a'\cdot a' = -m^2 a_0^2 \, e^{\phi}
    \;, \qquad
    \phi < 0
    \;.
\end{equation}
The individual components of $a'$ may contain oscillatory factors, c.f.~the brackets in~\eqref{eq:sin-sq-pulse}, without affecting~\eqref{rising-def}.
(Simple extensions are to continue directly to $\phi>0$ as a model of an ever-increasing field, or use $e^{-|\phi|}$ to model a sharply peaked field~\cite{Kibble:1965zza,Adorno:2016bjx}.) With this choice, we bring~\eqref{eq:h-2LL} to standard Abel form by changing the independent variable from $\phi$ to
\begin{equation}
    z := 1 -\delta \int\limits_{-\infty}^\phi\! \ud y\, a'(y) \cdot a'(y) = 1 + \Delta e^\phi
    \;,
\end{equation}
which, note, is just the solution for $h$ from \LLn{1}.
Our ODE~\eqref{eq:h-2LL} reduces to
\begin{equation}
    \label{solveme}
    h(z)\partial_z h(z) -h(z) = 2\delta \;.
\end{equation}
Setting all \emph{explicit} factors of $\delta\to 0$ in this expression recovers the \LLn{1} solution; this implies, as suggested in Ref.~\cite{DiPiazza:2018luu} that the difference between \LLn{1} and \LLn{2} depends essentially on $\delta$ (which should be small in the classical regime), rather than the potentially large $\Delta$.
To confirm this, we need the solution of~\eqref{solveme}.
This is easily found by first solving the equation for $z$ as a function of $h$.
Choosing initial condition $h(z=1)=1$, the solution is
\begin{equation}
     z = h - 2\delta \log \frac{h+2\delta}{1 + 2\delta}
     \;.
\end{equation}
Again, setting $\delta = 0$ on the right correctly recovers the \LLn{1} result: the difference is \emph{explicitly} dependent only on $\delta$.
Rearranging, we can write $h$ in terms of the Lambert function~\cite{Corless}, or product logarithm, $W$,
\begin{equation}
    \label{h-sol-W-1}
   -\frac{h(\phi)}{2\delta}
   =
   1 + W\bigg(
        -\frac{1+2\delta}{2\delta} \exp\bigg[
            -1 - \frac{1 + \Delta e^\phi}{2\delta}
        \bigg]
    \bigg)
    \;.
\end{equation}
Interestingly, the argument of $W$ is in the range $(-1/e, 0)$ which means we must take $W \equiv W_{-1}$, i.e.~\emph{not} the principal branch of $W$.
As is clear from the expressions above, and as shown explicitly in Fig.~\ref{FIG:PRODUCTLOG}, the difference between $h$ as predicted by \LLn{2} vs.\ \LLn{1} is extremely small unless the energy parameter $\delta$ is taken to be very large, and therefore outside the classical regime.

\subsection{Longitudinal polarisation}

For the longitudinally polarised field, \eqref{2LL-general} implies the following \LLn{2} equation for the transverse velocity components,
\begin{equation}
\label{deriv2}
    \dot{u}^\LCperp
    =
- 4 \tau_0 u^\LCp u^\LCm \left[ ( 1 + 2 \tau_0 u^\LCp \partial_\LCp ) (f_{\LCp \LCm})^2 \right]
    u^\LCperp
    \,
    .
\end{equation}
Comparing to~\eqref{eq:LL-perp}, the difference is the derivative term inside the brackets; this does not have a definite sign, and so the approach to the RFD is no longer uniform.
However, this new NLO term is roughly of order $\alpha \omega p^\LCp/m^2$ relative to the LO term, so is again dependent on energy rather than field strength.
As such, higher-order corrections are only significant when the energy parameter is large.
We conclude therefore that the RFD hypothesis continues to hold (at least) for parameters where a classical treatment is valid.

\section{Conclusions}
\label{sec:conclusion}

Exact solutions of equations of motion allow us to make statements that do not rely on approximations, and can give explicit insight into complex phenomena induced by, as considered here, radiation reaction.
We have presented new exact solutions of the Landau-Lifshitz equation, for a longitudinally polarised electric field depending on a single lightlike coordinate, and of the `second-order-of-reduction' Landau-Lifshitz equation (\LLn{2}) for a plane wave.

We have used these solutions to examine the radiation-free direction hypothesis~\cite{Gonoskov2018}, that is, the proposed universal approach of particle motion, in strong fields, to a direction which minimises radiation losses.
We have explicitly confirmed this behaviour in our new solutions.

Many authors have found exact solutions to the LAD and/or LL equations in a number of field configurations, including
fields depending only on time~\cite{RevModPhys.33.37},
electromagnetic fields that are constant~\cite{ares1999exact,Yaremko:2014qoa,PhysRevE.102.033210} or have a two-parameter symmetry group~\cite{Kazinski:2013vga},
rotating electric fields~\cite{bulanov2011lorentz,PhysRevE.66.046618,PhysRevE.102.033210},
the Coulomb potential in the non-relativistic limit~\cite{Rajeev:2008sw},
and plane waves~\cite{Heintzmann:1972mn,PiazzaExact}.
Many of these works contain implicit support for the RFD hypothesis.
For example, in a rotating electric field $\mathbf{E} = E_0 \left( \cos \omega t , \sin \omega t , 0 \right)$, the particle momentum co-rotates with the electric force, lagging $e\mathbf{E}$ by an angle $\approx 90^\circ$ for small $E_0$ and $\sim E_0^{-3/4}$ for large $E_0$
\cite[Sec.~II.A]{bulanov2011lorentz}.
Further, although the exact solution of LAD is not known even in a general constant field, its asymptotic behaviour is known~\cite{Kazinski:2010ce}, namely the particle worldline becomes confined to an eigen-2-plane of the field tensor;
the same holds for the exact solution of \LLn{1}~\cite{Yaremko:2014qoa}.

There are many examples of particle motion for which the Lorentz force law is integrable, or even superintegrable~\cite{Miller:2013gxa,Bagrov:2014rss,Heinzl:2017zsr}.
In future work it would be interesting to examine in detail how this integrability is affected by the addition of radiation-reaction terms in the LAD, LL and higher order, \LLn{n}, forms.
For example, the Lorentz force equation in a plane wave is superintegrable; in going to the  Landau-Lifshitz equation (\LLn{1}), one loses conserved quantities (i.e.~$\pi^\LCp$ is no longer conserved), which lowers the degree of integrability, while the second-order Landau-Lifshitz equation (\LLn{2}) becomes nonlinear and is only integrable in special cases.

Finding exact solutions of higher-reduction-of-order equations could shed light on the emergence of non-perturbative phenomena (runaways and acausal solutions), but this would require resummation~\cite{Zhang:2013ria}.
This is particularly intriguing as resummation has recently been highlighted as being essential for fully understanding the behaviour of \emph{quantum} dynamics in strong fields~\cite{Mironov:2020gbi,Edwards:2020npu,Heinzl:2021mji,Torgrimsson:2021wcj}.

\begin{acknowledgments}
\emph{The authors thank Ben King for useful comments and discussions.
The authors are supported by the Leverhulme Trust (RE, AI, TH), grant RPG-2019-148, and the EPSRC (AI), grant EP/S010319/1.}
\end{acknowledgments}

\bibliography{RFD-Bib}

\end{document}